\documentclass[prl,preprint,showpacs,showkeys,aps,floatfix]{revtex4-1}
\usepackage{amssymb,amsmath,amsthm,mathtools}
\usepackage{graphicx}
\usepackage{units}
\usepackage[usenames,dvipsnames,svgnames,table]{xcolor}

\newcommand{\pa}{\partial_1}
\newcommand{\pb}{\partial_2}
\newcommand{\pt}{\partial_t}
\newcommand{\ket}[1]{\left| #1 \right>}                                        
\newcommand{\bra}[1]{\left< #1 \right|}
\newcommand{\expect}[1]{\left< #1 \right>}                                     
\newcommand{\braket}[2]{\left< #1 \vphantom{#2} \right| \left. #2 \vphantom{#1} \right>}
\newcommand{\matrixel}[3]{\left< #1 \vphantom{#2#3} \right| #2 \left| #3 \vphantom{#1#2} \right>}

\begin{document}

\title{Exact single-electron approach to the dynamics of molecules in strong laser fields}
\author{Axel Schild}
\affiliation{Max-Planck-Institut f\"ur Mikrostrukturphysik}
\author{E.K.U. Gross}
\affiliation{Max-Planck-Institut f\"ur Mikrostrukturphysik}

\date{\today}

\begin{abstract}
  We present an exact single-electron picture that describes the correlated 
  electron dynamics in strong laser fields.
  Our approach is based on the factorization of the electronic wavefunction 
  as a product of a marginal and a conditional amplitude.
  The marginal amplitude, which depends only on one electronic coordinate 
  and yields the exact one-electron density and current density, obeys a 
  time-dependent Schr\"odinger equation with an effective time-dependent
  potential.
  The exact equations are used to derive an approximation that is a step 
  towards a general and feasible {\it ab-initio} single-electron approximation 
  for molecules.
  The derivation also challenges the usual interpretation of the single-active 
  electron approximation.
  From the study of model systems, we find that the exact and approximate 
  single-electron potentials for processes with negligible two-electron 
  ionization lead to a qualitatively similar dynamics, but that the ionization
  barrier may be explicitly time-dependent.
\end{abstract}
%

\maketitle

Molecules in strong laser fields are a fascinating field of research.
In such laser fields it becomes possible to monitor the electron dynamics in 
chemical reactions experimentally on its natural time scale, and in concert 
with the nuclear dynamics.
Recent experimental studies show that parts of the electron dynamics can 
already be measured.\cite{smirnova09,kraus15,wolter16,huppert16}

With progress in experimental techniques comes the necessity to develop 
and to improve theoretical tools to analyze the experiments.
An important role in this respect have single-electron pictures.
The 3-step model \cite{schafer93,corkum93} and its quantum version, the strong 
field approximation,\cite{lewenstein94} are single-electron models that describe 
the main mechanism which is responsible for many of the observed effects, 
e.g.\ high-harmonic generation.
Based on the success of these models, a Single-Active Electron (SAE) 
approximation \cite{kulander87} is often the basis for the development of 
quantum theories of strong field processes.\cite{awasthi08,le16}
From such investigations, general phenomena that may occur in experiments can 
be deduced.
However, there is no clear understanding of why the SAE works and what its 
limitations are, or even if it can be derived.\cite{rohringer06}
Thus, it is highly desirable to investigate how far we can get with a 
single-electron model.

Typically, in the SAE approximation a time-dependent Schr\"odinger equation
\begin{equation}
 i \pt \chi^{\mathtt{SAE}}(r_1,t) = \left(-\frac{\pa^2}{2} + \epsilon^{\mathtt{SAE}}(r_1) + r_1 F(t) \right) \chi^{\mathtt{SAE}}(r_1,t)
 \label{eq:sae}
\end{equation}
is solved for a single-electron wavefunction $\chi^{\mathtt{SAE}}$
and observables are computed from this wavefunction.\cite{rohringer06,le16}
The many-electron effects are approximated by an effective time-independent
potential $\epsilon^{\mathtt{SAE}}(r_1)$, while the interaction with the laser field 
may e.g.\ be described in the dipole approximation in the length gauge, as 
is done in \eqref{eq:sae}.
The crucial information for the SAE is the effective potential 
$\epsilon^{\mathtt{SAE}}$.
While for atoms, it may be possible to guess a model potential, 
this is much harder for molecules.\cite{awasthi08,barth08,abusamha10,awasthi10}
However, hints that more general model potentials can describe effects that 
seem to be beyond the applicability of the SAE approximation exists already 
for a while.\cite{watson97,lein05,gordon06}

In this article, we first present an exact single-electron description of 
a many-electron system in a laser field, the Exact Electron 
Factorization (EEF).
The EEF is then used to derive an approximation, the Time-Independent 
Conditional Amplitude (TICA) approximation, that is formally close to the SAE 
approximation \eqref{eq:sae} but has a different interpretation:
The SAE approximation assumes that the processes to be described 
are essentially single-electron processes and seems to treat all but one 
electron as ``frozen''.
Hence, it is often assumed that it cannot describe multi-electron 
effects.\cite{ishikawa15}
In contrast, the EEF and also the TICA approximation represent the dynamics of 
all electrons with an effective potential.

The EEF is a generalization of the Exact (Electron-Nuclear) Factorization,\cite{abedi10}
which separates the nuclear from the electronic system 
of a molecular wavefunction exactly, to the case of electrons only.
The idea of the EEF was already given for static systems some time 
ago\cite{hunter86,buijse89}, and aspects of it are also known in the field of 
Density Functional Theory.\cite{deb83,levy84,march86,march87}
With this article, we generalize the EEF to time-dependent problems and show 
that it is useful for developing the theory of attosecond experiments.
The derivation of a TICA approximation from the EEF shows the 
assumptions that are made when an equation such as \eqref{eq:sae} is used to 
represent the electron dynamics.
Also, it yields a general procedure to obtain the TICA potential for any system.

To make the ideas that follow as clear as possible, we write the general 
equations using a simplified notation with only two spatial coordinates, 
$r_1$ and $r_2$. 
In an $n$-electron system, $r_1$ and $r_2$ stand for an arbitrary 
partitioning of the coordinates of the electrons into two sets.
The most important case for our application is the case where $r_1$ 
contains the coordinates of $1$ electron and $r_2$ contains the 
coordinates of the remaining $n-1$ electrons.
This case is assumed below.
Also, we use atomic units and we do not use explicit vector 
notation.
The general equations with vector notation are given in the supplementary 
material.

We consider a non-relativistic description of a molecule in a laser field.
In the dipole approximation in the length gauge, the evolution of 
the system is described with the time-dependent Schr\"odinger equation
\begin{equation}
 i \pt \Psi = \left(-\frac{\pa^2}{2} -\frac{\pb^2}{2} + V(r_1,r_2) + (r_1 + r_2) F(t) \right) \Psi.
 \label{eq:tdse}
\end{equation}
Here, $V(r_1,r_2)$ is the Coulomb interaction among the electrons and of 
the electrons with clamped nuclei, and $F(t)$ is the time-dependent electric 
field.
We note that by using the reverse factorization \cite{suzuki14,khosravi15}, 
\eqref{eq:tdse} can be valid without clamping the nuclei.
However, then $V$ would be time-dependent and would not be a bare Coulomb 
potential.
The electronic wavefunction $\Psi(r_1,r_2;s_1,s_2|t)$ depends on spatial 
coordinates $r_j$ and spin coordinates $s_j$, which are in general not 
separable, but $\Psi(r_1,r_2;s_1,s_2|t)$ may always be written as a sum of 
coordinate permutations of a unique spatial wavefunction $\psi(r_1,r_2|t)$, 
multiplied by a corresponding spin function $\sigma(s_1,s_2)$.\cite{shpilkin96}
Below, we work with the spatial wavefunction $\psi(r_1,r_2|t)$ alone which,
for our problem, has the same information content as $\Psi(r_1,r_2;s_1,s_2|t)$
and which has a time-evolution given by the Schr\"odinger equation 
\eqref{eq:tdse}, too.

Next, we make the EEF ansatz
\begin{equation}
 \psi(r_1,r_2|t) = \chi(r_1|t) \phi(r_2|r_1,t)
 \label{eq:ansatz}
\end{equation}
with partial normalization condition
\begin{equation}
 \braket{\phi(r_2|r_1,t)}{\phi(r_2|r_1,t)}_2 \stackrel{!}{=} 1 ~~ \forall r_1, t,
\end{equation}
where the notation $\expect{\cdot}_2$ represents integration over all 
coordinates $r_2$.
It automatically follows that 
\begin{equation}
 |\chi(r_1|t)|^2 = \braket{\psi(r_1,r_2|t)}{\psi(r_1,r_2|t)}_2.
 \label{eq:rho}
\end{equation}
Given that $|\psi(r_1,r_2|t)|^2$ is normalized to one and hence has the meaning 
of a joint probability density, i.e.\ it represents the probability of finding 
one electron at $r_1$ and the other electrons at $r_2$ given we are at time $t$,
$\chi$ and $\phi$ also acquire a special meaning: 
$|\chi(r_1|t)|^2$ is the one-electron density or marginal density, i.e.\ it 
represents the probability of finding an electron at $r_1$, given time $t$;
$|\phi(r_2|r_1,t)|^2$ is the conditional probability of finding $n-1$ electrons 
at configuration $r_2$, given one electron is at $r_1$ and given time $t$.
Hence, we call $\chi$ the marginal amplitude and $\phi$ the conditional 
amplitude. 
We note that if the number of spin-up and spin-down electrons is not equal,
there are two different factorizations.
An example is the 3-electron system discussed below.

A variational derivation of the equations of motion for $\chi$ and $\phi$
yields 
\begin{equation}
 i \pt \chi = \left(\frac{(-i\pa+A(r_1,t))^2}{2} + \epsilon(r_1,t) \right) \chi
 \label{eq:eom_chi}
\end{equation}
for the marginal amplitude. 
This is a normal time-dependent Schr\"odinger equation with a vector potential
\begin{equation}
 A(r_1,t) = \operatorname{Im} \braket{\phi}{\pa \phi}_2
 \label{eq:eef_a}
\end{equation}
and a scalar potential
\begin{equation}
 \epsilon(r_1,t) = \epsilon_{\mathtt{T}} + \epsilon_{\mathtt{V}} + \epsilon_{\mathtt{F}} + \epsilon_{\mathtt{FS}} + \epsilon_{\mathtt{GD}}
 \label{eq:pes_ex}
\end{equation}
with average kinetic and potential energy of the other electrons
\begin{equation}
 \epsilon_{\mathtt{T}}(r_1,t) + \epsilon_{\mathtt{V}}(r_1,t) = \matrixel{\phi}{-\frac{\pb^2}{2}+V}{\phi}_2,
\end{equation}
the electric field interaction with a modified dipole operator
\begin{equation}
 \epsilon_{\mathtt{F}}(r_1,t) = F(t) (r_1 + \matrixel{\phi}{r_2}{\phi}_2),
\end{equation}
a Fubini-Study term
\begin{equation}
 \epsilon_{\mathtt{FS}}(r_1,t) = \frac{1}{2} \matrixel{\pa\phi}{\left(1-\ket{\phi}\bra{\phi}\right)}{\pa\phi}_2,
\end{equation}
and a gauge-dependent term 
\begin{equation}
 \epsilon_{\mathtt{GD}}(r_1,t) = \operatorname{Im} \braket{\phi}{\pt \phi}_2.
\end{equation}
There is a gauge freedom in the choice of a phase $S(r_1,t)$, because
$\tilde{\chi} = e^{-iS(r_1,t)} \chi$ and $\tilde{\phi} = e^{iS(r_1,t)} \phi$
yield the same electronic wavefunction according to \eqref{eq:ansatz},
and the equations of motion stay invariant up to the change $\tilde{A} = A + \pa S$,
$\tilde{\epsilon}_{\mathtt{GD}} = \epsilon_{\mathtt{GD}} + \pt S$.
The equation of motion for the conditional amplitude $\phi$ is not 
of interest here and can be found in the supplementary material.

The marginal amplitude $\chi(r_1|t)$ is an interesting quantity.
It yields the exact one-electron density, cf.\ \eqref{eq:rho},
and it obeys a time-dependent Schr\"odinger equation \eqref{eq:eom_chi}.
Additionally, it is straightforward to show that it also yields 
the exact one-electron current density.
From $\chi$, all observables depending on $r_1$ and $\pa$, most notably
the dipole expectation value (that yields, e.g., the high harmonic generation
spectrum) and the momentum expectation value, can be obtained.
Consequently, $\chi$ may also be called a one-electron wavefunction, and 
it gives an exact single-electron picture of the dynamics.

The marginal amplitude $\chi$ is an appealing quantity, but
solving the full problem does not by itself become easier by making 
ansatz \eqref{eq:ansatz}.
Instead, the main problem now is to obtain the scalar and vector potentials 
$\epsilon$ and $A$, which depend on the conditional amplitude $\phi$.
However, the single-electron Schr\"odinger equation \eqref{eq:eom_chi} 
gives us the possibility \emph{to derive a single-electron approximation}.

One way to derive a single-electron approximation of the form \eqref{eq:sae} 
from the exact single-electron equation \eqref{eq:eom_chi} is to assume that 
the conditional amplitude is time-independent, $\phi(r_2|r_1,t) \stackrel{!}{=} 
\phi_0(r_2|r_1)$.
Together with choosing the gauge as $A(r_1,t) = 0$ (which may not
always be possible\cite{requist16}), we obtain a time-independent potential 
\begin{equation}
 \epsilon^{\mathtt{TICA}}(r_1) = \epsilon_{\mathtt{T}}[\phi_0] + \epsilon_{\mathtt{V}}[\phi_0] + \epsilon_{\mathtt{FS}}[\phi_0].
 \label{eq:pes_sae}
\end{equation}
The only formal difference between this approximation, called the TICA
(Time-Independent Conditional Amplitude) approximation, and the SAE equation 
\eqref{eq:sae} is the change of the dipole 
operator from $r_1$ to $d(r_1) = r_1 + \matrixel{\phi_0}{r_2}{\phi_0}$, i.e. 
the TICA approximation is
\begin{equation}
  i \pt \chi^{\mathtt{TICA}} = \left(-\frac{\pa^2}{2} + \epsilon^{\mathtt{TICA}}(r_1) + d(r_1) F(t) \right) \chi^{\mathtt{TICA}}.
 \label{eq:sae_eef}
\end{equation}
A typical choice for $\phi_0$ is the conditional amplitude of the state at 
$t=0$, which usually is an eigenstate of the system.
Then, to compute $\epsilon^{\mathtt{TICA}}$ in practice it is only necessary to 
know the one-electron density $\rho(r_1)$ of this state,  as $\rho(r_1) = 
|\chi^{\mathtt{TICA}}|^2$ at $t=0$, and to solve the time-independent analogue 
of \eqref{eq:sae_eef} for $\epsilon^{\mathtt{TICA}}$.
The modified dipole $d(r_1)$ can be obtained by computing 
$\matrixel{\psi}{r_2}{\psi} / \rho_1$ for the initial wavefunction.
While in our examples studied below the function $d(r_1)$ is of minor 
importance, it can become important for many electrons.

A significant part of the computational effort of a TICA calculation 
is to obtain $\rho(r_1)$.
$\epsilon^{\mathtt{TICA}}(r_1)$ can then be computed for any system, in contrast to the 
SAE potential $\epsilon^{\mathtt{SAE}}(r_1)$, which is {\it a priori} unknown.
However, it is necessary to divide by $\rho(r_1)$ to compute both 
$\epsilon^{\mathtt{TICA}}$ and $d(r_1)$, which may be numerically difficult.
Then, to obtain the dynamics it is only necessary to solve the single-electron 
time-dependent Schr\"odinger equation \eqref{eq:sae_eef}, independent of the 
number of electrons in the system.

\begin{figure}[htbp]
  \centering
  \includegraphics[width=0.7\textwidth]{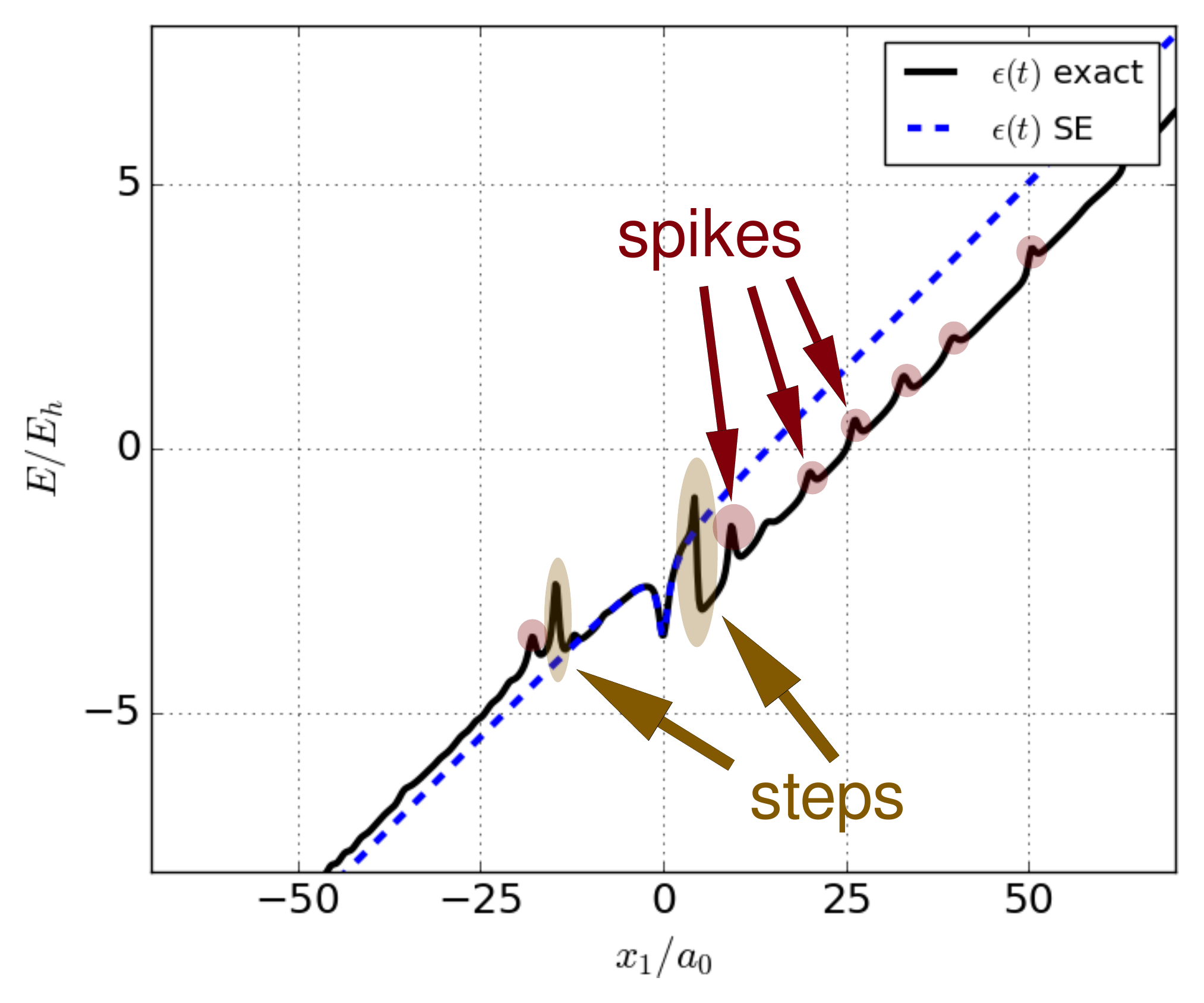}
  \caption{Exact single-electron potential \eqref{eq:pes_ex} (solid black line)
  and TICA potential \eqref{eq:pes_sae} with laser potential (broken blue line) 
  at a time where the amplitude of the laser field is maximal, for the 
  one-dimensional 2-electron model.}
  \label{fig:cmp_1d}
\end{figure}

To learn more about the exact and the TICA potential, we consider
one-dimensional models of the helium atom (2 electrons) and of the lithium atom
(3 electrons).
All solutions to the involved eigenvalue problems were obtained with help of 
the linear algebra routines in the SciPy package \cite{scipy,lehoucq98}.
The time-propagation was performed with the Gonoskov-Marklund 
propagator\cite{gonoskov16} or a Runge-Kutta method.
Details can be found in the supplementary material.

For the Helium model, we choose as initial state the spin-singlet ground state
and use the parameters of \cite{shilin13}.
Our choice for $\phi_0$ is the conditional amplitude of the initial state.
The dynamics is computed for a 12-cycle laser pulse with a wavelength of 
\unit[580]{nm} and a maximum intensity of \unit[$6.9 \times 10^{14}$]{W/cm$^2$}.
By comparing the dynamics and the high harmonic generation spectra,
we found quantitative agreement of the TICA calculations with the 
exact calculations, and the effect of the modified dipole operator 
is almost negligible.
A representative picture for the comparison of the exact single-electron 
potential $\epsilon(r_1,t)$ with the TICA potential is shown in 
figure \ref{fig:cmp_1d}, and a movie is given in the supplementary material.

We see two important features that distinguish the exact potential 
from the TICA potential:
There are time-dependent steps and spikes in the exact potential
which are absent in the TICA potential.
The steps are similar to those known from Time-Dependent Density Functional 
Theory \cite{lein05,elliot12,hodgson13,hodgson16} and are related to those 
from the Exact Electron Nuclear Factorization.\cite{abedi13}
They occur only at certain times and positions, develop and disappear rapidly,
and they can only be found in the gauge-dependent part $\epsilon_{\mathtt{GD}}$ of 
the potential.
We do not have an intuitive interpretation for the steps in the EEF, 
but we find that they have negligible effect on the dynamics:
They are located at positions with very low electron density and the parts of 
the potential connected by steps are parallel.
Hence, they change the momentum of some parts of $\chi$, but this does not 
result in an important effect for the overall dynamics.
The spikes also occur at positions where the one-electron density is minimal.
They are a peculiar feature of the factorization ansatz:
because $|\chi|^2$ is the one-electron density and this never becomes zero
in our systems,
we do not have exact nodes except if we impose them by a choice of the phase 
(which may lead to discontinuities in $\chi$).
We found that in the model propagation, the spikes in the exact potential can 
be neglected for the propagation of $\chi$, which is equivalent to allowing
the marginal amplitude to become zero.

Next, we study the lithium model, a one-dimensional spin-doublet model system 
with parameters taken from \cite{rapp14}, for an 8-cycle laser 
pulse with several laser frequencies between \unit[0.1]{$E_h\hbar$} (\unit[152]{nm}) 
and \unit[1.0]{$E_h\hbar$} (\unit[46]{nm}), and with a maximum intensity of 
\unit[$8.8 \times 10^{13}$]{W/cm$^2$}.
The electronic wavefunction is given by
\begin{equation}
 \Psi = N \left(\psi(r_1,r_2,r_3) \ket{\uparrow \uparrow \downarrow}  \right. \\ \left. + 
                \psi(r_2,r_3,r_1) \ket{\uparrow \downarrow \uparrow} +
                \psi(r_3,r_1,r_2) \ket{\downarrow \uparrow \uparrow} \right)
\end{equation}
with anti-symmetry condition $\psi(r_1,r_2,r_3) = -\psi(r_2,r_1,r_3)$
for the spatial wavefunction.
There are two possible factorizations, one for the spin-up one-electron
density 
$|\chi_{\uparrow}(r_1|t)|^2 = \braket{\psi(r_1,r_2,r_3)}{\psi(r_1,r_2,r_3)}_{23}$
and one for the spin-down one-electron density 
$|\chi_{\downarrow}(r_3|t)|^2 = \braket{\psi(r_1,r_2,r_3)}{\psi(r_1,r_2,r_3)}_{12}$.
As we aim at describing processes that mainly involve the valence electron, 
which is a spin-up electron, we only consider the factorization for 
$\chi_{\uparrow}$.

\begin{figure}[htbp]
  \centering
  \includegraphics[width=0.7\textwidth]{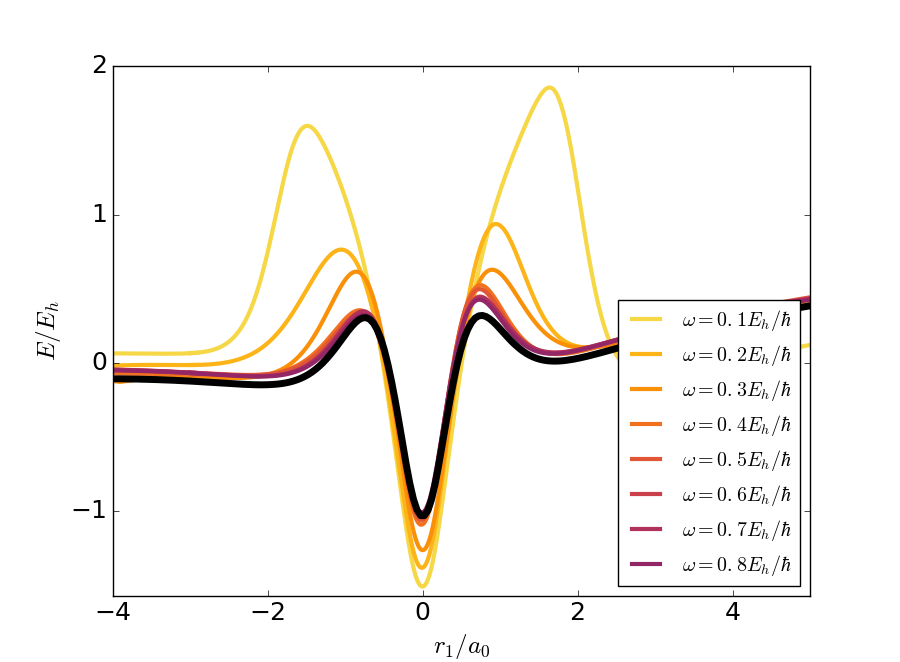}
  \caption{Exact single-electron potential \eqref{eq:pes_ex} (colored lines)
  for different values of the laser frequency, and TICA potential 
  \eqref{eq:pes_sae} with laser potential (black line) after half of the pulse,
  where the amplitude of the laser field is maximal, for the one-dimensional 
  3-electron model.}
  \label{fig:cmp_3d}
\end{figure}

Figure \ref{fig:cmp_3d} shows a snapshot of the exact and the TICA potential 
for the 3-electron model system in the laser fields, at a time where the 
electric field is maximal, in the spatial region close to the nucleus.
Movies of the dynamics for different frequencies of the laser field are
given in the supplementary material.
Note that the TICA potential is independent of the laser frequency, as it is a 
sum of the initial exact potential and the laser interaction in dipole form, 
cf.\ \eqref{eq:sae_eef}.

The initial potential shows a deep minimum, but also a barrier to the left and 
to the right followed by shallow minima.
This reflects the electronic structure of the problem.
It also illustrates that a suitable model potential for many-electron systems 
is not easy to guess and an {\it ab-inito} treatment like the TICA approximation 
is recommended.
During the evolution, time-dependent spikes and steps occur like in the 
2-electron case.
What is most striking, however, is the explicit time-dependence of the exact 
potential around the bounding region:
At times where the field is strong, the potential well located around 
$r_1=0$ becomes deeper, and the barriers at its sides become higher.
These changes become stronger with smaller frequency of the laser pulse.
Interpreted in terms of the 3-step model, the effect of this additional barrier 
is clear:
It suppresses the tunneling out of the bounding well and is an obstacle during 
the recombination step.
In contrast, the TICA potential does not show this time-dependent barrier,
which results in a similar qualitative dynamics, but different quantitative 
results.

Further investigations of the contributions to the exact potential show that 
this effect originates in the interaction of the two spin-up electrons, while 
the effect of the contributions of the spin-down electron to the exact potential 
is small.
We expect the time-dependent barrier to be partly a consequence of the 
anti-symmetry condition of the electronic wavefunction for the two spin-up 
electrons, which is not met by the TICA:
While the EEF is exact and preserves symmetries of the full 
wavefunction, it does not reflect those symmetries in the equations of motion.
Consequently, approximations to the EEF lead to a symmetry breaking, which
may be of minor importance, like in the 2-electron spin singlet case, but 
which we expect to be significant in most cases, like in the 3-electron spin 
doublet case.
Hence, it may be worthwhile to include symmetries of the problems explicitly.
Extensions like a repeated factorization\cite{cederbaum15} or more symmetric 
factorizations need to be explored in the future.

In summary, from our investigations we learned the following:
First, it is possible to obtain a single-electron approximation analogous to 
the usual SAE approximation, the TICA approximation, from an exact 
single-electron theory, the EEF.
It can be applied not only for atoms or simple molecules, but also 
for complex molecules where the alignment of the molecule relative to the laser 
field matters.
Additionally, the derivation from the EEF also highlights one of the less 
appreciated merits of some SAE calculations: 
They do not only describe an ``active'' electron and a ``frozen'' 
core, but they also describes the dynamics of electrons in the core region in an 
approximate way.
This becomes clear if the SAE wavefunction is interpreted like the TICA 
wavefunction, i.e.\ as an approximation to the exact single-electron density
and current density.
Second, we found that a 2-electron spin-singlet system behaves similar to a 
one-electron system and that the shape of the exact potential is simple to model,
at least as long as two-electron ionization is negligible. 
Hence it is easy to approximate, but its applicability as model for many-electron 
effects is limited.
Third, we saw that already for the 3-electron model system an {\it ab-initio} 
treatment like the TICA approximation is necessary to obtain a potential which 
includes all relevant features.
Finally, it became clear that the TICA approach
is useful to compute qualitative 
effects, again with the restriction of negligible two-electron ionization.
It does, however, in general ignore a time-dependent barrier that changes
the ionization and recombination step.
This barrier, in turn, challenges our view on the 3-step model and shows 
the need for further studies.
Also, the role of the vector potential \eqref{eq:eef_a} and its relation to 
topological features of the system \cite{requist16} needs to be investigated in
the future.

\bibliography{references}{}

\begin{thebibliography}{39}%
\makeatletter
\providecommand \@ifxundefined [1]{%
 \@ifx{#1\undefined}
}%
\providecommand \@ifnum [1]{%
 \ifnum #1\expandafter \@firstoftwo
 \else \expandafter \@secondoftwo
 \fi
}%
\providecommand \@ifx [1]{%
 \ifx #1\expandafter \@firstoftwo
 \else \expandafter \@secondoftwo
 \fi
}%
\providecommand \natexlab [1]{#1}%
\providecommand \enquote  [1]{``#1''}%
\providecommand \bibnamefont  [1]{#1}%
\providecommand \bibfnamefont [1]{#1}%
\providecommand \citenamefont [1]{#1}%
\providecommand \href@noop [0]{\@secondoftwo}%
\providecommand \href [0]{\begingroup \@sanitize@url \@href}%
\providecommand \@href[1]{\@@startlink{#1}\@@href}%
\providecommand \@@href[1]{\endgroup#1\@@endlink}%
\providecommand \@sanitize@url [0]{\catcode `\\12\catcode `\$12\catcode
  `\&12\catcode `\#12\catcode `\^12\catcode `\_12\catcode `\%12\relax}%
\providecommand \@@startlink[1]{}%
\providecommand \@@endlink[0]{}%
\providecommand \url  [0]{\begingroup\@sanitize@url \@url }%
\providecommand \@url [1]{\endgroup\@href {#1}{\urlprefix }}%
\providecommand \urlprefix  [0]{URL }%
\providecommand \Eprint [0]{\href }%
\providecommand \doibase [0]{http://dx.doi.org/}%
\providecommand \selectlanguage [0]{\@gobble}%
\providecommand \bibinfo  [0]{\@secondoftwo}%
\providecommand \bibfield  [0]{\@secondoftwo}%
\providecommand \translation [1]{[#1]}%
\providecommand \BibitemOpen [0]{}%
\providecommand \bibitemStop [0]{}%
\providecommand \bibitemNoStop [0]{.\EOS\space}%
\providecommand \EOS [0]{\spacefactor3000\relax}%
\providecommand \BibitemShut  [1]{\csname bibitem#1\endcsname}%
\let\auto@bib@innerbib\@empty
\bibitem [{\citenamefont {Smirnova}\ \emph {et~al.}(2009)\citenamefont
  {Smirnova}, \citenamefont {Mairesse}, \citenamefont {Patchkovskii},
  \citenamefont {Dudovich}, \citenamefont {Villeneuve}, \citenamefont
  {Corkum},\ and\ \citenamefont {Ivanov}}]{smirnova09}%
  \BibitemOpen
  \bibfield  {author} {\bibinfo {author} {\bibfnamefont {O.}~\bibnamefont
  {Smirnova}}, \bibinfo {author} {\bibfnamefont {Y.}~\bibnamefont {Mairesse}},
  \bibinfo {author} {\bibfnamefont {S.}~\bibnamefont {Patchkovskii}}, \bibinfo
  {author} {\bibfnamefont {N.}~\bibnamefont {Dudovich}}, \bibinfo {author}
  {\bibfnamefont {D.}~\bibnamefont {Villeneuve}}, \bibinfo {author}
  {\bibfnamefont {P.}~\bibnamefont {Corkum}}, \ and\ \bibinfo {author}
  {\bibfnamefont {M.~Y.}\ \bibnamefont {Ivanov}},\ }\href {\doibase
  10.1038/nature08253} {\bibfield  {journal} {\bibinfo  {journal} {Nature}\
  }\textbf {\bibinfo {volume} {460}},\ \bibinfo {pages} {972} (\bibinfo {year}
  {2009})}\BibitemShut {NoStop}%
\bibitem [{\citenamefont {Kraus}\ \emph {et~al.}(2015)\citenamefont {Kraus},
  \citenamefont {Mignolet}, \citenamefont {Baykusheva}, \citenamefont
  {Rupenyan}, \citenamefont {Horn{\'y}}, \citenamefont {Penka}, \citenamefont
  {Grassi}, \citenamefont {Tolstikhin}, \citenamefont {Schneider},
  \citenamefont {Jensen}, \citenamefont {Madsen}, \citenamefont {Bandrauk},
  \citenamefont {Remacle},\ and\ \citenamefont {W{\"o}rner}}]{kraus15}%
  \BibitemOpen
  \bibfield  {author} {\bibinfo {author} {\bibfnamefont {P.~M.}\ \bibnamefont
  {Kraus}}, \bibinfo {author} {\bibfnamefont {B.}~\bibnamefont {Mignolet}},
  \bibinfo {author} {\bibfnamefont {D.}~\bibnamefont {Baykusheva}}, \bibinfo
  {author} {\bibfnamefont {A.}~\bibnamefont {Rupenyan}}, \bibinfo {author}
  {\bibfnamefont {L.}~\bibnamefont {Horn{\'y}}}, \bibinfo {author}
  {\bibfnamefont {E.~F.}\ \bibnamefont {Penka}}, \bibinfo {author}
  {\bibfnamefont {G.}~\bibnamefont {Grassi}}, \bibinfo {author} {\bibfnamefont
  {O.~I.}\ \bibnamefont {Tolstikhin}}, \bibinfo {author} {\bibfnamefont
  {J.}~\bibnamefont {Schneider}}, \bibinfo {author} {\bibfnamefont
  {F.}~\bibnamefont {Jensen}}, \bibinfo {author} {\bibfnamefont {L.~B.}\
  \bibnamefont {Madsen}}, \bibinfo {author} {\bibfnamefont {A.~D.}\
  \bibnamefont {Bandrauk}}, \bibinfo {author} {\bibfnamefont {F.}~\bibnamefont
  {Remacle}}, \ and\ \bibinfo {author} {\bibfnamefont {H.~J.}\ \bibnamefont
  {W{\"o}rner}},\ }\href {\doibase 10.1126/science.aab2160} {\bibfield
  {journal} {\bibinfo  {journal} {Science}\ }\textbf {\bibinfo {volume}
  {350}},\ \bibinfo {pages} {790} (\bibinfo {year} {2015})}\BibitemShut
  {NoStop}%
\bibitem [{\citenamefont {Wolter}\ \emph {et~al.}(2016)\citenamefont {Wolter},
  \citenamefont {Pullen}, \citenamefont {Le}, \citenamefont {Baudisch},
  \citenamefont {Doblhoff-Dier}, \citenamefont {Senftleben}, \citenamefont
  {Hemmer}, \citenamefont {Schr{\"o}ter}, \citenamefont {Ullrich},
  \citenamefont {Pfeifer}, \citenamefont {Moshammer}, \citenamefont
  {Gr{\"a}fe}, \citenamefont {Vendrell}, \citenamefont {Lin},\ and\
  \citenamefont {Biegert}}]{wolter16}%
  \BibitemOpen
  \bibfield  {author} {\bibinfo {author} {\bibfnamefont {B.}~\bibnamefont
  {Wolter}}, \bibinfo {author} {\bibfnamefont {M.~G.}\ \bibnamefont {Pullen}},
  \bibinfo {author} {\bibfnamefont {A.-T.}\ \bibnamefont {Le}}, \bibinfo
  {author} {\bibfnamefont {M.}~\bibnamefont {Baudisch}}, \bibinfo {author}
  {\bibfnamefont {K.}~\bibnamefont {Doblhoff-Dier}}, \bibinfo {author}
  {\bibfnamefont {A.}~\bibnamefont {Senftleben}}, \bibinfo {author}
  {\bibfnamefont {M.}~\bibnamefont {Hemmer}}, \bibinfo {author} {\bibfnamefont
  {C.~D.}\ \bibnamefont {Schr{\"o}ter}}, \bibinfo {author} {\bibfnamefont
  {J.}~\bibnamefont {Ullrich}}, \bibinfo {author} {\bibfnamefont
  {T.}~\bibnamefont {Pfeifer}}, \bibinfo {author} {\bibfnamefont
  {R.}~\bibnamefont {Moshammer}}, \bibinfo {author} {\bibfnamefont
  {S.}~\bibnamefont {Gr{\"a}fe}}, \bibinfo {author} {\bibfnamefont
  {O.}~\bibnamefont {Vendrell}}, \bibinfo {author} {\bibfnamefont {C.~D.}\
  \bibnamefont {Lin}}, \ and\ \bibinfo {author} {\bibfnamefont
  {J.}~\bibnamefont {Biegert}},\ }\href {\doibase 10.1126/science.aah3429}
  {\bibfield  {journal} {\bibinfo  {journal} {Science}\ }\textbf {\bibinfo
  {volume} {354}},\ \bibinfo {pages} {308} (\bibinfo {year}
  {2016})}\BibitemShut {NoStop}%
\bibitem [{\citenamefont {Huppert}\ \emph {et~al.}(2016)\citenamefont
  {Huppert}, \citenamefont {Jordan}, \citenamefont {Baykusheva}, \citenamefont
  {von Conta},\ and\ \citenamefont {W\"orner}}]{huppert16}%
  \BibitemOpen
  \bibfield  {author} {\bibinfo {author} {\bibfnamefont {M.}~\bibnamefont
  {Huppert}}, \bibinfo {author} {\bibfnamefont {I.}~\bibnamefont {Jordan}},
  \bibinfo {author} {\bibfnamefont {D.}~\bibnamefont {Baykusheva}}, \bibinfo
  {author} {\bibfnamefont {A.}~\bibnamefont {von Conta}}, \ and\ \bibinfo
  {author} {\bibfnamefont {H.~J.}\ \bibnamefont {W\"orner}},\ }\href {\doibase
  10.1103/PhysRevLett.117.093001} {\bibfield  {journal} {\bibinfo  {journal}
  {Phys. Rev. Lett.}\ }\textbf {\bibinfo {volume} {117}},\ \bibinfo {pages}
  {093001} (\bibinfo {year} {2016})}\BibitemShut {NoStop}%
\bibitem [{\citenamefont {Schafer}\ \emph {et~al.}(1993)\citenamefont
  {Schafer}, \citenamefont {Yang}, \citenamefont {DiMauro},\ and\ \citenamefont
  {Kulander}}]{schafer93}%
  \BibitemOpen
  \bibfield  {author} {\bibinfo {author} {\bibfnamefont {K.~J.}\ \bibnamefont
  {Schafer}}, \bibinfo {author} {\bibfnamefont {B.}~\bibnamefont {Yang}},
  \bibinfo {author} {\bibfnamefont {L.~F.}\ \bibnamefont {DiMauro}}, \ and\
  \bibinfo {author} {\bibfnamefont {K.~C.}\ \bibnamefont {Kulander}},\ }\href
  {\doibase 10.1103/PhysRevLett.70.1599} {\bibfield  {journal} {\bibinfo
  {journal} {Phys. Rev. Lett.}\ }\textbf {\bibinfo {volume} {70}},\ \bibinfo
  {pages} {1599} (\bibinfo {year} {1993})}\BibitemShut {NoStop}%
\bibitem [{\citenamefont {Corkum}(1993)}]{corkum93}%
  \BibitemOpen
  \bibfield  {author} {\bibinfo {author} {\bibfnamefont {P.~B.}\ \bibnamefont
  {Corkum}},\ }\href {\doibase 10.1103/PhysRevLett.71.1994} {\bibfield
  {journal} {\bibinfo  {journal} {Phys. Rev. Lett.}\ }\textbf {\bibinfo
  {volume} {71}},\ \bibinfo {pages} {1994} (\bibinfo {year}
  {1993})}\BibitemShut {NoStop}%
\bibitem [{\citenamefont {Lewenstein}\ \emph {et~al.}(1994)\citenamefont
  {Lewenstein}, \citenamefont {Balcou}, \citenamefont {Ivanov}, \citenamefont
  {L'Huillier},\ and\ \citenamefont {Corkum}}]{lewenstein94}%
  \BibitemOpen
  \bibfield  {author} {\bibinfo {author} {\bibfnamefont {M.}~\bibnamefont
  {Lewenstein}}, \bibinfo {author} {\bibfnamefont {P.}~\bibnamefont {Balcou}},
  \bibinfo {author} {\bibfnamefont {M.~Y.}\ \bibnamefont {Ivanov}}, \bibinfo
  {author} {\bibfnamefont {A.}~\bibnamefont {L'Huillier}}, \ and\ \bibinfo
  {author} {\bibfnamefont {P.~B.}\ \bibnamefont {Corkum}},\ }\href {\doibase
  10.1103/PhysRevA.49.2117} {\bibfield  {journal} {\bibinfo  {journal} {Phys.
  Rev. A}\ }\textbf {\bibinfo {volume} {49}},\ \bibinfo {pages} {2117}
  (\bibinfo {year} {1994})}\BibitemShut {NoStop}%
\bibitem [{\citenamefont {Kulander}(1987)}]{kulander87}%
  \BibitemOpen
  \bibfield  {author} {\bibinfo {author} {\bibfnamefont {K.~C.}\ \bibnamefont
  {Kulander}},\ }\href {\doibase 10.1103/PhysRevA.35.445} {\bibfield  {journal}
  {\bibinfo  {journal} {Phys. Rev. A}\ }\textbf {\bibinfo {volume} {35}},\
  \bibinfo {pages} {445} (\bibinfo {year} {1987})}\BibitemShut {NoStop}%
\bibitem [{\citenamefont {Awasthi}\ \emph {et~al.}(2008)\citenamefont
  {Awasthi}, \citenamefont {Vanne}, \citenamefont {Saenz}, \citenamefont
  {Castro},\ and\ \citenamefont {Decleva}}]{awasthi08}%
  \BibitemOpen
  \bibfield  {author} {\bibinfo {author} {\bibfnamefont {M.}~\bibnamefont
  {Awasthi}}, \bibinfo {author} {\bibfnamefont {Y.~V.}\ \bibnamefont {Vanne}},
  \bibinfo {author} {\bibfnamefont {A.}~\bibnamefont {Saenz}}, \bibinfo
  {author} {\bibfnamefont {A.}~\bibnamefont {Castro}}, \ and\ \bibinfo {author}
  {\bibfnamefont {P.}~\bibnamefont {Decleva}},\ }\href {\doibase
  10.1103/PhysRevA.77.063403} {\bibfield  {journal} {\bibinfo  {journal} {Phys.
  Rev. A}\ }\textbf {\bibinfo {volume} {77}},\ \bibinfo {pages} {063403}
  (\bibinfo {year} {2008})}\BibitemShut {NoStop}%
\bibitem [{\citenamefont {Le}\ \emph {et~al.}(2016)\citenamefont {Le},
  \citenamefont {Wei}, \citenamefont {Jin},\ and\ \citenamefont {Lin}}]{le16}%
  \BibitemOpen
  \bibfield  {author} {\bibinfo {author} {\bibfnamefont {A.-T.}\ \bibnamefont
  {Le}}, \bibinfo {author} {\bibfnamefont {H.}~\bibnamefont {Wei}}, \bibinfo
  {author} {\bibfnamefont {C.}~\bibnamefont {Jin}}, \ and\ \bibinfo {author}
  {\bibfnamefont {C.~D.}\ \bibnamefont {Lin}},\ }\href {\doibase
  10.1088/0953-4075/49/5/053001} {\bibfield  {journal} {\bibinfo  {journal} {J.
  Phys. B: At. Mol. Opt. Phys.}\ }\textbf {\bibinfo {volume} {49}},\ \bibinfo
  {pages} {053001} (\bibinfo {year} {2016})}\BibitemShut {NoStop}%
\bibitem [{\citenamefont {Rohringer}\ \emph {et~al.}(2006)\citenamefont
  {Rohringer}, \citenamefont {Gordon},\ and\ \citenamefont
  {Santra}}]{rohringer06}%
  \BibitemOpen
  \bibfield  {author} {\bibinfo {author} {\bibfnamefont {N.}~\bibnamefont
  {Rohringer}}, \bibinfo {author} {\bibfnamefont {A.}~\bibnamefont {Gordon}}, \
  and\ \bibinfo {author} {\bibfnamefont {R.}~\bibnamefont {Santra}},\ }\href
  {\doibase 10.1103/PhysRevA.74.043420} {\bibfield  {journal} {\bibinfo
  {journal} {Phys. Rev. A}\ }\textbf {\bibinfo {volume} {74}},\ \bibinfo
  {pages} {043420} (\bibinfo {year} {2006})}\BibitemShut {NoStop}%
\bibitem [{\citenamefont {Barth}\ \emph {et~al.}(2008)\citenamefont {Barth},
  \citenamefont {Manz},\ and\ \citenamefont {Paramonov}}]{barth08}%
  \BibitemOpen
  \bibfield  {author} {\bibinfo {author} {\bibfnamefont {I.}~\bibnamefont
  {Barth}}, \bibinfo {author} {\bibfnamefont {J.}~\bibnamefont {Manz}}, \ and\
  \bibinfo {author} {\bibfnamefont {G.}~\bibnamefont {Paramonov}},\ }\href
  {\doibase 10.1080/00268970701871007} {\bibfield  {journal} {\bibinfo
  {journal} {Mol. Phys.}\ }\textbf {\bibinfo {volume} {106}},\ \bibinfo {pages}
  {467} (\bibinfo {year} {2008})}\BibitemShut {NoStop}%
\bibitem [{\citenamefont {Abu-samha}\ and\ \citenamefont
  {Madsen}(2010)}]{abusamha10}%
  \BibitemOpen
  \bibfield  {author} {\bibinfo {author} {\bibfnamefont {M.}~\bibnamefont
  {Abu-samha}}\ and\ \bibinfo {author} {\bibfnamefont {L.~B.}\ \bibnamefont
  {Madsen}},\ }\href {\doibase 10.1103/PhysRevA.81.033416} {\bibfield
  {journal} {\bibinfo  {journal} {Phys. Rev. A}\ }\textbf {\bibinfo {volume}
  {81}},\ \bibinfo {pages} {033416} (\bibinfo {year} {2010})}\BibitemShut
  {NoStop}%
\bibitem [{\citenamefont {Awasthi}\ and\ \citenamefont
  {Saenz}(2010)}]{awasthi10}%
  \BibitemOpen
  \bibfield  {author} {\bibinfo {author} {\bibfnamefont {M.}~\bibnamefont
  {Awasthi}}\ and\ \bibinfo {author} {\bibfnamefont {A.}~\bibnamefont
  {Saenz}},\ }\href {\doibase 10.1103/PhysRevA.81.063406} {\bibfield  {journal}
  {\bibinfo  {journal} {Phys. Rev. A}\ }\textbf {\bibinfo {volume} {81}},\
  \bibinfo {pages} {063406} (\bibinfo {year} {2010})}\BibitemShut {NoStop}%
\bibitem [{\citenamefont {Watson}\ \emph {et~al.}(1997)\citenamefont {Watson},
  \citenamefont {Sanpera}, \citenamefont {Lappas}, \citenamefont {Knight},\
  and\ \citenamefont {Burnett}}]{watson97}%
  \BibitemOpen
  \bibfield  {author} {\bibinfo {author} {\bibfnamefont {J.~B.}\ \bibnamefont
  {Watson}}, \bibinfo {author} {\bibfnamefont {A.}~\bibnamefont {Sanpera}},
  \bibinfo {author} {\bibfnamefont {D.~G.}\ \bibnamefont {Lappas}}, \bibinfo
  {author} {\bibfnamefont {P.~L.}\ \bibnamefont {Knight}}, \ and\ \bibinfo
  {author} {\bibfnamefont {K.}~\bibnamefont {Burnett}},\ }\href
  {https://dx.doi.org/10.1103/PhysRevLett.78.1884} {\bibfield  {journal}
  {\bibinfo  {journal} {Phys. Rev. Lett.}\ }\textbf {\bibinfo {volume} {78}},\
  \bibinfo {pages} {1884} (\bibinfo {year} {1997})}\BibitemShut {NoStop}%
\bibitem [{\citenamefont {Lein}\ and\ \citenamefont {K\"ummel}(2005)}]{lein05}%
  \BibitemOpen
  \bibfield  {author} {\bibinfo {author} {\bibfnamefont {M.}~\bibnamefont
  {Lein}}\ and\ \bibinfo {author} {\bibfnamefont {S.}~\bibnamefont
  {K\"ummel}},\ }\href {\doibase 10.1103/PhysRevLett.94.143003} {\bibfield
  {journal} {\bibinfo  {journal} {Phys. Rev. Lett.}\ }\textbf {\bibinfo
  {volume} {94}},\ \bibinfo {pages} {143003} (\bibinfo {year}
  {2005})}\BibitemShut {NoStop}%
\bibitem [{\citenamefont {Gordon}\ \emph {et~al.}(2006)\citenamefont {Gordon},
  \citenamefont {K\"artner}, \citenamefont {Rohringer},\ and\ \citenamefont
  {Santra}}]{gordon06}%
  \BibitemOpen
  \bibfield  {author} {\bibinfo {author} {\bibfnamefont {A.}~\bibnamefont
  {Gordon}}, \bibinfo {author} {\bibfnamefont {F.~X.}\ \bibnamefont
  {K\"artner}}, \bibinfo {author} {\bibfnamefont {N.}~\bibnamefont
  {Rohringer}}, \ and\ \bibinfo {author} {\bibfnamefont {R.}~\bibnamefont
  {Santra}},\ }\href {\doibase 10.1103/PhysRevLett.96.223902} {\bibfield
  {journal} {\bibinfo  {journal} {Phys. Rev. Lett.}\ }\textbf {\bibinfo
  {volume} {96}},\ \bibinfo {pages} {223902} (\bibinfo {year}
  {2006})}\BibitemShut {NoStop}%
\bibitem [{\citenamefont {Ishikawa}\ and\ \citenamefont
  {Sato}(2015)}]{ishikawa15}%
  \BibitemOpen
  \bibfield  {author} {\bibinfo {author} {\bibfnamefont {K.~L.}\ \bibnamefont
  {Ishikawa}}\ and\ \bibinfo {author} {\bibfnamefont {T.}~\bibnamefont
  {Sato}},\ }\href {\doibase 10.1109/JSTQE.2015.2438827} {\bibfield  {journal}
  {\bibinfo  {journal} {IEEE Journal of Selected Topics in Quantum
  Electronics}\ }\textbf {\bibinfo {volume} {21}},\ \bibinfo {pages} {1}
  (\bibinfo {year} {2015})}\BibitemShut {NoStop}%
\bibitem [{\citenamefont {Abedi}\ \emph {et~al.}(2010)\citenamefont {Abedi},
  \citenamefont {Maitra},\ and\ \citenamefont {Gross}}]{abedi10}%
  \BibitemOpen
  \bibfield  {author} {\bibinfo {author} {\bibfnamefont {A.}~\bibnamefont
  {Abedi}}, \bibinfo {author} {\bibfnamefont {N.~T.}\ \bibnamefont {Maitra}}, \
  and\ \bibinfo {author} {\bibfnamefont {E.~K.~U.}\ \bibnamefont {Gross}},\
  }\href {\doibase 10.1103/PhysRevLett.105.123002} {\bibfield  {journal}
  {\bibinfo  {journal} {Phys. Rev. Lett.}\ }\textbf {\bibinfo {volume} {105}},\
  \bibinfo {pages} {123002} (\bibinfo {year} {2010})}\BibitemShut {NoStop}%
\bibitem [{\citenamefont {Hunter}(1986)}]{hunter86}%
  \BibitemOpen
  \bibfield  {author} {\bibinfo {author} {\bibfnamefont {G.}~\bibnamefont
  {Hunter}},\ }\href {\doibase 10.1002/qua.560290209} {\bibfield  {journal}
  {\bibinfo  {journal} {Int. J. Quant. Chem.}\ }\textbf {\bibinfo {volume}
  {29}},\ \bibinfo {pages} {197} (\bibinfo {year} {1986})}\BibitemShut
  {NoStop}%
\bibitem [{\citenamefont {A.~Buijse}\ \emph {et~al.}(1989)\citenamefont
  {A.~Buijse}, \citenamefont {Baerends},\ and\ \citenamefont
  {Snijders}}]{buijse89}%
  \BibitemOpen
  \bibfield  {author} {\bibinfo {author} {\bibfnamefont {M.}~\bibnamefont
  {A.~Buijse}}, \bibinfo {author} {\bibfnamefont {E.~J.}\ \bibnamefont
  {Baerends}}, \ and\ \bibinfo {author} {\bibfnamefont {J.~G.}\ \bibnamefont
  {Snijders}},\ }\href {\doibase 10.1103/PhysRevA.40.4190} {\bibfield
  {journal} {\bibinfo  {journal} {Phys. Rev. A}\ }\textbf {\bibinfo {volume}
  {40}},\ \bibinfo {pages} {4190} (\bibinfo {year} {1989})}\BibitemShut
  {NoStop}%
\bibitem [{\citenamefont {Deb}\ and\ \citenamefont {Ghosh}(1983)}]{deb83}%
  \BibitemOpen
  \bibfield  {author} {\bibinfo {author} {\bibfnamefont {B.~M.}\ \bibnamefont
  {Deb}}\ and\ \bibinfo {author} {\bibfnamefont {S.~K.}\ \bibnamefont
  {Ghosh}},\ }\href {\doibase 10.1002/qua.560230104} {\bibfield  {journal}
  {\bibinfo  {journal} {International Journal of Quantum Chemistry}\ }\textbf
  {\bibinfo {volume} {23}},\ \bibinfo {pages} {1} (\bibinfo {year}
  {1983})}\BibitemShut {NoStop}%
\bibitem [{\citenamefont {Levy}\ \emph {et~al.}(1984)\citenamefont {Levy},
  \citenamefont {Perdew},\ and\ \citenamefont {Sahni}}]{levy84}%
  \BibitemOpen
  \bibfield  {author} {\bibinfo {author} {\bibfnamefont {M.}~\bibnamefont
  {Levy}}, \bibinfo {author} {\bibfnamefont {J.~P.}\ \bibnamefont {Perdew}}, \
  and\ \bibinfo {author} {\bibfnamefont {V.}~\bibnamefont {Sahni}},\ }\href
  {\doibase 10.1103/PhysRevA.30.2745} {\bibfield  {journal} {\bibinfo
  {journal} {Phys. Rev. A}\ }\textbf {\bibinfo {volume} {30}},\ \bibinfo
  {pages} {2745} (\bibinfo {year} {1984})}\BibitemShut {NoStop}%
\bibitem [{\citenamefont {March}(1986)}]{march86}%
  \BibitemOpen
  \bibfield  {author} {\bibinfo {author} {\bibfnamefont {N.}~\bibnamefont
  {March}},\ }\href {\doibase http://dx.doi.org/10.1016/0375-9601(86)90123-4}
  {\bibfield  {journal} {\bibinfo  {journal} {Phys. Lett. A}\ }\textbf
  {\bibinfo {volume} {113}},\ \bibinfo {pages} {476 } (\bibinfo {year}
  {1986})}\BibitemShut {NoStop}%
\bibitem [{\citenamefont {March}(1987)}]{march87}%
  \BibitemOpen
  \bibfield  {author} {\bibinfo {author} {\bibfnamefont {N.~H.}\ \bibnamefont
  {March}},\ }\href {\doibase 10.1002/jcc.540080414} {\bibfield  {journal}
  {\bibinfo  {journal} {J. Comp. Chem.}\ }\textbf {\bibinfo {volume} {8}},\
  \bibinfo {pages} {375} (\bibinfo {year} {1987})}\BibitemShut {NoStop}%
\bibitem [{\citenamefont {Suzuki}\ \emph {et~al.}(2014)\citenamefont {Suzuki},
  \citenamefont {Abedi}, \citenamefont {Maitra}, \citenamefont {Yamashita},\
  and\ \citenamefont {Gross}}]{suzuki14}%
  \BibitemOpen
  \bibfield  {author} {\bibinfo {author} {\bibfnamefont {Y.}~\bibnamefont
  {Suzuki}}, \bibinfo {author} {\bibfnamefont {A.}~\bibnamefont {Abedi}},
  \bibinfo {author} {\bibfnamefont {N.~T.}\ \bibnamefont {Maitra}}, \bibinfo
  {author} {\bibfnamefont {K.}~\bibnamefont {Yamashita}}, \ and\ \bibinfo
  {author} {\bibfnamefont {E.~K.~U.}\ \bibnamefont {Gross}},\ }\href {\doibase
  10.1103/PhysRevA.89.040501} {\bibfield  {journal} {\bibinfo  {journal} {Phys.
  Rev. A}\ }\textbf {\bibinfo {volume} {89}},\ \bibinfo {pages} {040501}
  (\bibinfo {year} {2014})}\BibitemShut {NoStop}%
\bibitem [{\citenamefont {Khosravi}\ \emph {et~al.}(2015)\citenamefont
  {Khosravi}, \citenamefont {Abedi},\ and\ \citenamefont
  {Maitra}}]{khosravi15}%
  \BibitemOpen
  \bibfield  {author} {\bibinfo {author} {\bibfnamefont {E.}~\bibnamefont
  {Khosravi}}, \bibinfo {author} {\bibfnamefont {A.}~\bibnamefont {Abedi}}, \
  and\ \bibinfo {author} {\bibfnamefont {N.~T.}\ \bibnamefont {Maitra}},\
  }\href {\doibase 10.1103/PhysRevLett.115.263002} {\bibfield  {journal}
  {\bibinfo  {journal} {Phys. Rev. Lett.}\ }\textbf {\bibinfo {volume} {115}},\
  \bibinfo {pages} {263002} (\bibinfo {year} {2015})}\BibitemShut {NoStop}%
\bibitem [{\citenamefont {Shpilkin}\ \emph {et~al.}(1996)\citenamefont
  {Shpilkin}, \citenamefont {Smolenskii},\ and\ \citenamefont
  {Zefirov}}]{shpilkin96}%
  \BibitemOpen
  \bibfield  {author} {\bibinfo {author} {\bibfnamefont {S.~A.}\ \bibnamefont
  {Shpilkin}}, \bibinfo {author} {\bibfnamefont {E.~A.}\ \bibnamefont
  {Smolenskii}}, \ and\ \bibinfo {author} {\bibfnamefont {N.~S.}\ \bibnamefont
  {Zefirov}},\ }\href {\doibase 10.1021/ci950101u} {\bibfield  {journal}
  {\bibinfo  {journal} {J. Chem. Inf. Comput. Sci.}\ }\textbf {\bibinfo
  {volume} {36}},\ \bibinfo {pages} {409} (\bibinfo {year} {1996})}\BibitemShut
  {NoStop}%
\bibitem [{\citenamefont {Requist}\ \emph {et~al.}(2016)\citenamefont
  {Requist}, \citenamefont {Tandetzky},\ and\ \citenamefont
  {Gross}}]{requist16}%
  \BibitemOpen
  \bibfield  {author} {\bibinfo {author} {\bibfnamefont {R.}~\bibnamefont
  {Requist}}, \bibinfo {author} {\bibfnamefont {F.}~\bibnamefont {Tandetzky}},
  \ and\ \bibinfo {author} {\bibfnamefont {E.~K.~U.}\ \bibnamefont {Gross}},\
  }\href {\doibase 10.1103/PhysRevA.93.042108} {\bibfield  {journal} {\bibinfo
  {journal} {Phys. Rev. A}\ }\textbf {\bibinfo {volume} {93}},\ \bibinfo
  {pages} {042108} (\bibinfo {year} {2016})}\BibitemShut {NoStop}%
\bibitem [{\citenamefont {Jones}\ \emph {et~al.}(01  )\citenamefont {Jones},
  \citenamefont {Oliphant}, \citenamefont {Peterson} \emph {et~al.}}]{scipy}%
  \BibitemOpen
  \bibfield  {author} {\bibinfo {author} {\bibfnamefont {E.}~\bibnamefont
  {Jones}}, \bibinfo {author} {\bibfnamefont {T.}~\bibnamefont {Oliphant}},
  \bibinfo {author} {\bibfnamefont {P.}~\bibnamefont {Peterson}},  \emph
  {et~al.},\ }\href@noop {} {\enquote {\bibinfo {title} {{SciPy}: Open source
  scientific tools for {Python} (www.scipy.org)},}\ } (\bibinfo {year}
  {2001--})\BibitemShut {NoStop}%
\bibitem [{\citenamefont {Lehoucq}\ \emph {et~al.}(1998)\citenamefont
  {Lehoucq}, \citenamefont {Sorensen},\ and\ \citenamefont {Yang}}]{lehoucq98}%
  \BibitemOpen
  \bibfield  {author} {\bibinfo {author} {\bibfnamefont {R.~B.}\ \bibnamefont
  {Lehoucq}}, \bibinfo {author} {\bibfnamefont {D.~C.}\ \bibnamefont
  {Sorensen}}, \ and\ \bibinfo {author} {\bibfnamefont {C.}~\bibnamefont
  {Yang}},\ }\href@noop {} {\emph {\bibinfo {title} {ARPACK USERS GUIDE:
  Solution of Large Scale Eigenvalue Problems by Implicitly Restarted Arnoldi
  Methods}}}\ (\bibinfo  {publisher} {SIAM},\ \bibinfo {address} {Philadelphia,
  PA},\ \bibinfo {year} {1998})\BibitemShut {NoStop}%
\bibitem [{\citenamefont {Gonoskov}\ and\ \citenamefont
  {Marklund}(2016)}]{gonoskov16}%
  \BibitemOpen
  \bibfield  {author} {\bibinfo {author} {\bibfnamefont {I.}~\bibnamefont
  {Gonoskov}}\ and\ \bibinfo {author} {\bibfnamefont {M.}~\bibnamefont
  {Marklund}},\ }\href {\doibase http://dx.doi.org/10.1016/j.cpc.2016.02.006}
  {\bibfield  {journal} {\bibinfo  {journal} {Computer Physics Communications}\
  }\textbf {\bibinfo {volume} {202}},\ \bibinfo {pages} {211 } (\bibinfo {year}
  {2016})}\BibitemShut {NoStop}%
\bibitem [{\citenamefont {Shi-Lin}\ and\ \citenamefont
  {Ting-Yun}(2013)}]{shilin13}%
  \BibitemOpen
  \bibfield  {author} {\bibinfo {author} {\bibfnamefont {H.}~\bibnamefont
  {Shi-Lin}}\ and\ \bibinfo {author} {\bibfnamefont {S.}~\bibnamefont
  {Ting-Yun}},\ }\href {http://stacks.iop.org/1674-1056/22/i=1/a=013101}
  {\bibfield  {journal} {\bibinfo  {journal} {Chinese Physics B}\ }\textbf
  {\bibinfo {volume} {22}},\ \bibinfo {pages} {013101} (\bibinfo {year}
  {2013})}\BibitemShut {NoStop}%
\bibitem [{\citenamefont {Elliott}\ \emph {et~al.}(2012)\citenamefont
  {Elliott}, \citenamefont {Fuks}, \citenamefont {Rubio},\ and\ \citenamefont
  {Maitra}}]{elliot12}%
  \BibitemOpen
  \bibfield  {author} {\bibinfo {author} {\bibfnamefont {P.}~\bibnamefont
  {Elliott}}, \bibinfo {author} {\bibfnamefont {J.~I.}\ \bibnamefont {Fuks}},
  \bibinfo {author} {\bibfnamefont {A.}~\bibnamefont {Rubio}}, \ and\ \bibinfo
  {author} {\bibfnamefont {N.~T.}\ \bibnamefont {Maitra}},\ }\href {\doibase
  10.1103/PhysRevLett.109.266404} {\bibfield  {journal} {\bibinfo  {journal}
  {Phys. Rev. Lett.}\ }\textbf {\bibinfo {volume} {109}},\ \bibinfo {pages}
  {266404} (\bibinfo {year} {2012})}\BibitemShut {NoStop}%
\bibitem [{\citenamefont {Hodgson}\ \emph {et~al.}(2013)\citenamefont
  {Hodgson}, \citenamefont {Ramsden}, \citenamefont {Chapman}, \citenamefont
  {Lillystone},\ and\ \citenamefont {Godby}}]{hodgson13}%
  \BibitemOpen
  \bibfield  {author} {\bibinfo {author} {\bibfnamefont {M.~J.~P.}\
  \bibnamefont {Hodgson}}, \bibinfo {author} {\bibfnamefont {J.~D.}\
  \bibnamefont {Ramsden}}, \bibinfo {author} {\bibfnamefont {J.~B.~J.}\
  \bibnamefont {Chapman}}, \bibinfo {author} {\bibfnamefont {P.}~\bibnamefont
  {Lillystone}}, \ and\ \bibinfo {author} {\bibfnamefont {R.~W.}\ \bibnamefont
  {Godby}},\ }\href {\doibase 10.1103/PhysRevB.88.241102} {\bibfield  {journal}
  {\bibinfo  {journal} {Phys. Rev. B}\ }\textbf {\bibinfo {volume} {88}},\
  \bibinfo {pages} {241102} (\bibinfo {year} {2013})}\BibitemShut {NoStop}%
\bibitem [{\citenamefont {Hodgson}\ \emph {et~al.}(2016)\citenamefont
  {Hodgson}, \citenamefont {Ramsden},\ and\ \citenamefont {Godby}}]{hodgson16}%
  \BibitemOpen
  \bibfield  {author} {\bibinfo {author} {\bibfnamefont {M.~J.~P.}\
  \bibnamefont {Hodgson}}, \bibinfo {author} {\bibfnamefont {J.~D.}\
  \bibnamefont {Ramsden}}, \ and\ \bibinfo {author} {\bibfnamefont {R.~W.}\
  \bibnamefont {Godby}},\ }\href {\doibase 10.1103/PhysRevB.93.155146}
  {\bibfield  {journal} {\bibinfo  {journal} {Phys. Rev. B}\ }\textbf {\bibinfo
  {volume} {93}},\ \bibinfo {pages} {155146} (\bibinfo {year}
  {2016})}\BibitemShut {NoStop}%
\bibitem [{\citenamefont {Abedi}\ \emph {et~al.}(2013)\citenamefont {Abedi},
  \citenamefont {Agostini}, \citenamefont {Suzuki},\ and\ \citenamefont
  {Gross}}]{abedi13}%
  \BibitemOpen
  \bibfield  {author} {\bibinfo {author} {\bibfnamefont {A.}~\bibnamefont
  {Abedi}}, \bibinfo {author} {\bibfnamefont {F.}~\bibnamefont {Agostini}},
  \bibinfo {author} {\bibfnamefont {Y.}~\bibnamefont {Suzuki}}, \ and\ \bibinfo
  {author} {\bibfnamefont {E.~K.~U.}\ \bibnamefont {Gross}},\ }\href {\doibase
  10.1103/PhysRevLett.110.263001} {\bibfield  {journal} {\bibinfo  {journal}
  {Phys. Rev. Lett.}\ }\textbf {\bibinfo {volume} {110}},\ \bibinfo {pages}
  {263001} (\bibinfo {year} {2013})}\BibitemShut {NoStop}%
\bibitem [{\citenamefont {Rapp}\ and\ \citenamefont {Bauer}(2014)}]{rapp14}%
  \BibitemOpen
  \bibfield  {author} {\bibinfo {author} {\bibfnamefont {J.}~\bibnamefont
  {Rapp}}\ and\ \bibinfo {author} {\bibfnamefont {D.}~\bibnamefont {Bauer}},\
  }\href {\doibase 10.1103/PhysRevA.89.033401} {\bibfield  {journal} {\bibinfo
  {journal} {Phys. Rev. A}\ }\textbf {\bibinfo {volume} {89}},\ \bibinfo
  {pages} {033401} (\bibinfo {year} {2014})}\BibitemShut {NoStop}%
\bibitem [{\citenamefont {Cederbaum}(2015)}]{cederbaum15}%
  \BibitemOpen
  \bibfield  {author} {\bibinfo {author} {\bibfnamefont {L.~S.}\ \bibnamefont
  {Cederbaum}},\ }\href {\doibase 10.1016/j.chemphys.2015.05.021} {\bibfield
  {journal} {\bibinfo  {journal} {Chem. Phys.}\ }\textbf {\bibinfo {volume}
  {457}},\ \bibinfo {pages} {129} (\bibinfo {year} {2015})}\BibitemShut
  {NoStop}%
\end{thebibliography}%

\end{document}